%% file: paper.tex
\documentclass[conference]{IEEEtran}
\IEEEoverridecommandlockouts
\usepackage{cite}
\usepackage{amsmath,amssymb,amsfonts}
\usepackage{algorithmic}
\usepackage{graphicx}
\usepackage{textcomp}
\usepackage{xcolor}
\usepackage{hyperref}
\usepackage[nolist]{acronym} 
\usepackage{etoolbox}
\usepackage[absolute]{textpos}

\newcommand{\WBF}{\widehat{\diamond}}

\makeatletter
\patchcmd{\@makecaption}
  {\scshape}
  {}
  {}
  {}
\makeatletter
\patchcmd{\@makecaption}
  {\\}
  {.\ }
  {}
  {}
\makeatother
\def\tablename{Table}


\input{macrosIEEE}

\input{acro_list}

\def\BibTeX{{\rm B\kern-.05em{\sc i\kern-.025em b}\kern-.08em
    T\kern-.1667em\lower.7ex\hbox{E}\kern-.125emX}}
\begin{document}

\begin{textblock}{20}(-2,.25)
\noindent \centering This work is to be published in the Proceedings of \\ 2023 IEEE 22nd International Conference on Trust, Security and Privacy in Computing and Communications (TrustCom)
\end{textblock}

\title{Zero Trust Score-based Network-level \\ Access Control in Enterprise Networks
\thanks{© 2023 IEEE.  Personal use of this material is permitted.  Permission from IEEE must be obtained for all other uses, in any current or future media, including reprinting/republishing this material for advertising or promotional purposes, creating new collective works, for resale or redistribution to servers or lists, or reuse of any copyrighted component of this work in other works.

This work was supported by the bwNET2020+ project which is funded by the Ministry of Science, Research and the Arts Baden-Württemberg (MWK). The authors alone are responsible for the content of this paper.}
}

\author{
    \IEEEauthorblockN{
        Leonard~Bradatsch\IEEEauthorrefmark{2},
        Oleksandr~Miroshkin\IEEEauthorrefmark{3},
        Nataša~Trkulja\IEEEauthorrefmark{2},
        Frank~Kargl\IEEEauthorrefmark{2},
	}
    \IEEEauthorblockA{\IEEEauthorrefmark{2}
        Ulm~University,
        Institute~of~Distributed~Systems,
        Ulm,
        Germany\\
    }
    \IEEEauthorblockA{\IEEEauthorrefmark{3}
        Ulm~University,
        KIZ,
        Ulm,
        Germany\\
        Email:
        \{%
            leonard.bradatsch,%
            oleksandr.miroshkin,%
            natasa.trkulja,%
            frank.kargl\}@uni-ulm.de\\
    }
}

\maketitle

\input{chapters/00-abstract.tex}


\input{chapters/01-introduction}

\input{chapters/02-background}
\input{chapters/03-threat-model}
\input{chapters/04-threshold-calculation.tex}
\input{chapters/06-score-based-trust-algorithms.tex}
\input{chapters/07-evaluation-discussion.tex}
\input{chapters/08-conclusion}

\bibliographystyle{IEEEtranS}
\bibliography{sfc_literature.bib}

\end{document}

%% file: macrosIEEE.tex
\newlength{\figsize}
\newlength{\subfigwidth}
\newlength{\subfiglabelwidth}


%% file: acro_list.tex
\begin{acronym} 
\acro{SFC}{Service Function Chaining}
\acro{SF}{Service Function}
\acro{NSH}{Network Service Header}
\acro{SPI}{Service Path Identifier}
\acro{SI}{Service Index}
\acro{DPI}{Deep Packet Inspection}
\acro{SFF}{Service Function Forwarder}
\acro{SR}{Segment Routing}
\acro{ZT}{Zero Trust}
\acro{PoC}{Proof of Concept}
\acro{POT}{Proof of Transit}
\acro{INT}{In-band Network Telemetry}
\acro{COTS}{commercial off-the-shelf}
\acro{MITM}{Man in the Middle}
\acro{OSM}{Open Source MANO}
\acro{PSK}{pre-shared key}
\acro{PFS}{Perfect Forward Secrecy}
\end{acronym}

%% file: chapters/00-abstract.tex
\begin{abstract}

Zero Trust security has recently gained attention in enterprise network security. One of its key ideas is making network-level access decisions based on trust scores. However, score-based access control in the enterprise domain still lacks essential elements in our understanding, and in this paper, we contribute with respect to three crucial aspects. First, we provide a comprehensive list of 29 trust attributes that can be used to calculate a trust score. By introducing a novel mathematical approach, we demonstrate how to quantify these attributes. Second, we describe a dynamic risk-based method to calculate the trust threshold the trust score must meet for permitted access. Third, we introduce a novel trust algorithm based on Subjective Logic that incorporates the first two contributions and offers fine-grained decision possibilities. We discuss how this algorithm shows a higher expressiveness compared to a lightweight additive trust algorithm. Performance-wise, a prototype of the Subjective Logic-based approach showed similar calculation times for making an access decision as the additive approach. In addition, the dynamic threshold calculation showed only 7\% increased decision-making times compared to a static threshold.

\end{abstract}






%% file: chapters/01-introduction.tex
\section{Introduction}
\label{sec:introduction}
Recently, Zero Trust (ZT) security has gained popularity in enterprise network security, replacing traditional perimeter security, which has shortcomings like insider threats~\cite{kindervag1}. ZT enforces strict security requirements focused on resource protection and unauthorized access prevention~\cite{BrKa21, RoBo20}. This paper focuses on ZT's trust-based access control at the network level. It governs whether an entity, such as example employee Alice, can access network resources such as her enterprise's Gitlab code repository~\cite{garbisEnterpriseGuide}. Each resource access request (RAR) undergoes a trust-based access decision, performed by a trust algorithm~\cite{RoBo20}.

NIST divides trust algorithms into criteria-based and score-based algorithms~\cite{RoBo20}.
Criteria-based solutions such as XACML are proven and well-defined.
Therefore, they are not considered further in this paper.

In score-based trust algorithms, a trust score is calculated for each resource access request (RAR). This score reflects the trustworthiness of the entity, like Alice, based on trust attributes such as expected access time or correct password. These trust attributes are weighted to indicate their impact on the trust score. For instance, if Alice accesses at her usual time, the score rises by the trust attribute's weight. The more the trust attribute contributes to the trust in the entity, the higher its weight. Conversely, the trust score may drop for unusual access times. The RAR is approved if it exceeds a predefined threshold; otherwise, it is not. Score-based algorithms, which allow attribute weighting, offer more dynamic decisions than criteria-based ones~\cite{RoBo20}, making them a current research focus.

However, there is still a lack of consensus on important points in the trust score calculation in the enterprise domain~\cite{GiBa17}.
From our perspective, this circumstance includes the following three fundamental aspects:
First, the literature lacks a comprehensive list of trust attributes from the enterprise domain that can be considered for trust score calculation.
So far, only subsets of possible trust attributes are presented such as in~\cite{GiBa17, RoBo20} and~\cite{garbisEnterpriseGuide}.
Second, a mathematical model that allows the quantification of trust attribute weights is missing.
Most of existing literature such as~\cite{channel, chuan2020implementation, mehraj2020establishing} use fixed weights without further definition.
Third, the literature does not sufficiently address how to determine the threshold the trust score is compared with.
In works such as~\cite{ghate2021advanced, chuan2020implementation}, the threshold is not defined but only a static value. 

This lack motivates the main contributions of this paper:
\begin{itemize}
    \item A comprehensive list of trust attributes, based on a systematic literature review, for evaluating entity trust in an enterprise context. Additionally, we introduce a mathematical model for quantifying the corresponding attributes' weights.
    \item An approach for dynamic calculation of the threshold which we refer to as the risk level.
    \item A novel trust-score calculation method based on Subjective Logic [16], incorporating the above contributions and discussed regarding its pros and cons. Additionally, we offer a publicly accessible proof of concept evaluated for performance.
\end{itemize}

We also highlight the need for trust in three entities: \emph{user}, \emph{device}, and \emph{communication channel}. Additionally, we introduce a basic additive trust algorithm for comparison and performance evaluation against more complex methods like the Subjective-Logic-based approach.


The paper is organized as follows: Section~\ref{sec:background} provides the technical background, while Section~\ref{sec:threatModelTrustEntities} motivates the need for trust in the mentioned three entities. Section~\ref{sec:weightedAttributes} provides a comprehensive list of trust attributes and their mathematical quantification. Section~\ref{sec:thresholdCalculation} introduces a new risk level calculation method. A straightforward additive approach for illustrating these concepts is presented in Section~\ref{sec:additiveTrustAlgorithms}. Section~\ref{sec:slTrustAlgorithm} introduces a novel fine-grained trust algorithm using Subjective Logic. Section~\ref{sec:discussion} compares this with the additive approach, followed by a performance evaluation in Section~\ref{sec:evaluation}. Lastly, we conclude by outlining future work and summarizing in Section~\ref{sec:conclAndFutureWork}.

%% file: chapters/02-background.tex
\section{Background}
\label{sec:background}
\begin{figure}
    \centering
    \includegraphics[width=.77\columnwidth]{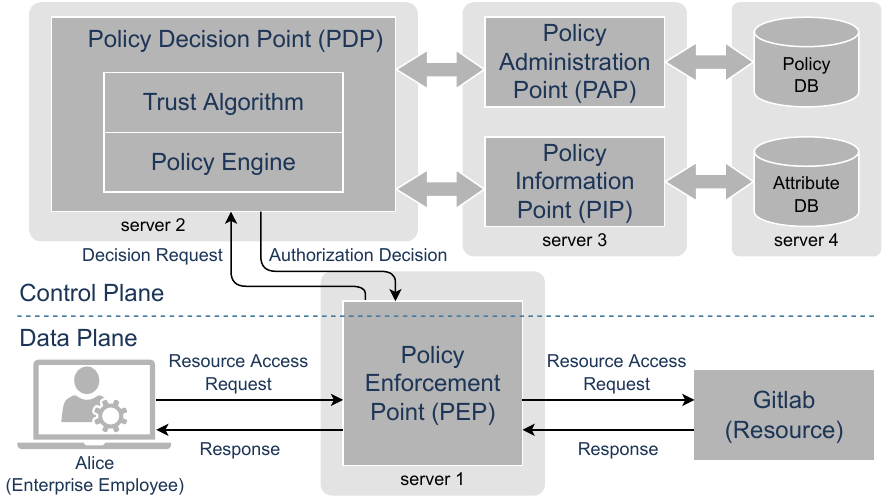}
    \vspace{-2mm}
    \caption{Decision-making process in a ZT network, adapted from NIST~\cite{RoBo20}}
    \label{fig:zt_background}
\end{figure}

This section describes how ZT enterprise architectures enforce score-based access decisions at the network level, following NIST~\cite{RoBo20} guidelines and depicted in Figure~\ref{fig:zt_background}. 

When an entity like Alice requests access to a resource, like her enterprise's Gitlab server, the Policy Enforcement Point (PEP) initiates a trust-based decision request for the respective RAR. The PEP forwards this decision request to the Policy Decision Point (PDP), where the Policy Engine (PE) processes it. For this purpose, the PE fetches the relevant score-based policy from the policy database via the Policy Administration Point (PAP). Policies are expressed in score-based enabled policy languages like UCON+~\cite{DiDi20} and are selected based on the requesting entity (here: Alice), action (here: resource access), requested resource (here: Gitlab) and context (e.g., RAR's date and time). Each policy specifies a trust algorithm and a threshold that represents the minimum amount of trust required for access. The specified trust algorithm calculates the trust score for the RAR, i.e., the entity involved in the RAR. This calculation uses trust attributes provided by a Policy Information Point (PIP). For example, Alice's usual access time could be a trust attribute. If the RAR is performed at Alice’s usual access time, the trust attribute is considered met, which leads to a trust score adjustment based on the weight of the met trust attribute. For an additive trust algorithm, for example, the weight is then added to the trust score. After evaluating all trust attributes, the trust algorithm compares the calculated trust score to the threshold. If met, access is granted; otherwise, it's denied. The PDP then communicates this access decision to the PEP, which enforces it.

%% file: chapters/03-threat-model.tex
\section{Trust Entities}
\label{sec:threatModelTrustEntities}
As previously outlined, the trust score is calculated for the entity, like Alice, initiating a RAR. 
However, such requests often involve more than just human entities; they can also include non-human entities, such as Alice's laptop~\cite{garbisEnterpriseGuide}. 
Defining the set of entities involved in a RAR is important, yet there's no consensus in existing literature on which entities should be considered.
In~\cite{RoBo20} and~\cite{mehraj2020establishing} user, application, and device are used as entities.
Related works such as~\cite{vanickis, garbisEnterpriseGuide, DiDi20, da2021zero, hatakeyama2021zero} and~\cite{BaOs16} consider the user and device as entities.
In the works~\cite{puthal2017building},~\cite{yao2020dynamic},~\cite{yang2022research} and~\cite{tao2018fine} the user as an entity is considered.
For the same purpose, a formal requestor is used as the entity in~\cite{xiaoning2012formal}.
In~\cite{chuan2020implementation, ramezanpour2021intelligent}, and~\cite{albuali2020ztimm}, the requesting hardware is used as the entity.
Zaheer et al.~\cite{zaheer2019eztrust} uses the microservices hosted on a server as the trust entity.
In~\cite{ghate2021advanced} and~\cite{mandal2021cloud}, the authors focus on attributes only.
Tian et al.~\cite{channel} uses the entities user, terminal, file, and channel.

We have focused on selecting a well-motivated set of entities from those already used in the literature. 
Based on the ZT threat model, we motivate the choice of three specific entities for the trust score calculation. 

Gilman et al. define a ZT threat model in which all communication channels and end systems, comprising both the user and the device, are presumed compromised~\cite{GiBa17}. For instance, the communication channel could have been eavesdropped, the user might be a rogue employee or the device could be malware-infected. Therefore, initial trust can not be assumed in the entities \emph{user}, \emph{device}, or \emph{communication channel}. 
Trust must be established in each individual entity, as a compromised entity can already cause damage on its own.

Individual trust scores can be calculated for each entity and summed first before comparison to a threshold, or each individual trust score can be compared to separate thresholds.
Sections~\ref{sec:additiveTrustAlgorithms} and~\ref{sec:slTrustAlgorithm} consider both a single total trust score and three separate scores, respectively. Section~\ref{sec:discussion} discusses the pros and cons of each method.

\section{Trust Attributes \& Trust Scores}
\label{sec:weightedAttributes}
As discussed before, score-based trust algorithms need to calculate trust scores for the entities \emph{user}, \emph{device}, and \emph{communication channel}.
This calculation is based on trust attributes.
A trust attribute is a property of an entity such as Alice's usual access times or her correct password that can be leveraged to build trust in that entity.
In different literature such as in~\cite{GiBa17, RoBo20, garbisEnterpriseGuide, sateesh2020state, BaOs16}, subsets of trust attributes are used. 
To the best of our knowledge, there is no comprehensive list of trust attributes in the context of enterprise networks.
Through systematic literature review, we collected all possible trust attributes and filtered them according to their applicability in the enterprise domain.
In the following, we provide the resulting list of trust attributes that are categorized based on the entity to which they apply:

\paragraph{User Trust Attributes}
\begin{itemize}
    \item Authentication Factors: Splits into one attribute per factor presented by the user, e.g., Password, Face ID, etc.~\cite{GiBa17}.
    \item Authentication Patterns: Indicates the user's typical authentication behavior, such as commonly used authentication factors~\cite{GiBa17}.
    \item Enterprise Presence: Reflects the time since the user last interacted with a resource and whether interaction is expected. For instance, interaction during the user's vacation would be suspicious~\cite{BaOs16}.
    \item Trust History: Records the trust scores from the user's previous RARs~\cite{GiBa17}.
    \item Input Behavior: Reflects user's usual input behavior. For example, it can be looked at how fast the user types~\cite{ali2017keystroke}.
    \item Service Usage: Considers whether the requested service is common for the user~\cite{RoBo20, GiBa17}.
    \item Device Usage:  Indicates if the device used for the RAR is among the user's known devices~\cite{RoBo20}.
    \item Access Time: Reflects if the RAR is made at a time that is common for that user~\cite{RoBo20}.
    \item Access Rate: Looks at whether the current request rate is within the usual range for this user~\cite{RoBo20, GiBa17}.
    \item Database Date Update: It is checked how up-to-date the data about the user is in the database~\cite{GiBa17}. 
\end{itemize}

\paragraph{Device Trust Attributes}
\begin{itemize}
    \item Connection Security: Reflects the willingness of the device to communicate securely. If not, it could indicate an attack, e.g., downgrade attack~\cite{sateesh2020state}. 
    \item Software Patch Level: Represents the security level of the software used for making the RAR~\cite{GiBa17}.
    \item System Patch Level: Reflects the security level of the system software, e.g., OS running on the device~\cite{garbisEnterpriseGuide}.
    \item Type: Considers the device's type or hardware configuration. For instance, personal computers are more prone to infection than smartphones~\cite{GiBa17, BaOs16}.
    \item Fingerprint: Depicts the device's unique software and hardware combination and checks if it aligns with the usual configuration~\cite{GiBa17}.
    \item Setup Date: Addresses the time elapsed since the device was last reset, with longer periods increasing the risk of infection~\cite{GiBa17}.
    \item Location: Checks if the device's current location matches one of its usual locations~\cite{RoBo20, GiBa17}.
    \item Health: Indicates the current hardware condition, such as CPU load ~\cite{GiBa17, garbisEnterpriseGuide}.
    \item Managed Device: Determines if the device is managed, serving as a basis for the relevance of other device attributes~\cite{RoBo20, BaOs16, GiBa17}.
    \item Vulnerability Scan: Results of the last antivirus scan~\cite{garbisEnterpriseGuide}.
    \item Authentication Factors, Enterprise Presence, Trust History, Service Usage, User Usage, Database Data Update: Analogous to the corresponding user attributes but in relation to the device.
    
\end{itemize}

\paragraph{Communication Channel (CC) Trust Attributes}
\begin{itemize}
    \item Authentication: Method of authentication for the CC.
    \item Confidentiality: Type of CC's confidentiality protection.
    \item Integrity: The way of integrity protection for the CC.
\end{itemize}

If not all attributes apply to a specific enterprise, a subset can be used with an accordingly adjusted threshold.

To be able to calculate a trust score, a trust attribute has a weight associated to it where this weight represents the proportionate impact on the increase of trust on this entity if this attribute is met. 
Additionally, there can be different degrees of fulfillment. 
For each degree of fulfillment, a separate weight can be defined.
For example, the weight can decrease the more the access time for Alice's RAR deviates from her usual access times, i.e. the more conspicuous it is.

In grey literature on ZT access control~\cite{GiBa17,RoBo20, BaOs16}, a quantifiable model for weighting trust attributes is is left open. 
In~\cite{ghate2021advanced}, the authors link trust attributes to the binary access decisions allow or deny.
Yao et al.~\cite{yang2022research} map trust attributes to three possible values: trusted, partially trusted, and untrusted.
In other academic literature ~\cite{channel, chuan2020implementation, mehraj2020establishing, tao2018fine, albuali2020ztimm, DiDi20} fixed weights are used without further definitions. To the best of our knowledge, there is currently no general model for quantifying trust attribute weights.

This lack motivates the following mathematical model that allows the quantification of trust attribute weights.
We start by defining four components for a trust attribute: $W_{tv}$, $W_{wv}$, $w_{av}$ and a weighting function $f_{W}$.
The set of target values $W_{tv}$ represent the trust attribute's values that lead to an increase in trust.
Dependent on the trust attribute, the possible target values can be, for example, the continuous set of real numbers or a discrete set of strings.
For the trust attribute evaluation, the trust in the entity is increased if $w_{av} \in W_{tv}$. 
$w_{av}$ is the actual value of this trust attribute at the time when the trust algorithm evaluates it. 
To determine the trust increment for the respective trust attribute's target values, a set of weight values $W_{wv}$ is defined for each trust attribute. 
Each target value in $W_{tv}$ is mapped onto a weight value in $W_{wv}$ so that $f_{W} : W_{tv} \rightarrow W_{wv}$. 
As described at the beginning of this section, the weighting function $f_{W}$ can be used to map different degrees of fulfillment, represented by the different elements of $W_{tv}$, to different weight values in $W_{wv}$. 
Values for all four components vary by enterprise and should be set by security experts. Examples are given in Sections~\ref{sec:additiveTrustAlgorithms} and~\ref{sec:slTrustAlgorithm}.

%% file: chapters/04-threshold-calculation.tex
\section{Risk Attributes \& Risk Level}
\label{sec:thresholdCalculation}
After the trust scores are calculated based on the trust attribute weights, they are compared to a predefined threshold. 
The threshold represents the minimum level of trust necessary to access the requested 
resource. 
In~\cite{channel} and~\cite{tao2018fine}, the necessary trust depends on the sensitivity of the requested resource. 
Vanickis et al.~\cite{vanickis} and Ramenzanpour et al.~\cite{ramezanpour2021intelligent} base the threshold level on the risk inherent in the access. 
In~\cite{ghate2021advanced, chuan2020implementation, mehraj2020establishing, xiaoning2012formal, ahmed2020protection, yang2022research, DiDi20, hatakeyama2021zero, GiBa17, RoBo20} and~\cite{albuali2020ztimm} the threshold is not defined. 
Yao et al.~\cite{yao2020dynamic} make their trust threshold dependent on the environment in which the resource is located.
In~\cite{da2021zero}, Da Silva et al. describe an context-dependent additive approach to calculate a security level akin to a threshold. However, their work focuses on smart homes and has limited applicability to enterprise networks.

To our knowledge, there is no approach to dynamically calculate the threshold in ZT enterprise networks in a clearly defined manner.
In the context of ZT, a relatively static value is used as a threshold without a clear explanation for how it was determined. 
We introduce an approach for dynamic threshold calculation based on a risk level.
A dynamic calculation allows a timely threshold adjustment according to changing risk conditions such as an increased network threat level.

We argue that the threshold depends on the risk inherent in accessing the requested resource. 
Risk is composed of the possible damage to the enterprise (impact) and the probability of damage (chance of occurrence)~\cite{shi2007advances}. 
The higher the level of risk, the higher the level of trust necessary, i.e., the higher the threshold.
Therefore, the threshold is equal to the risk level, and for the rest of the paper, we use the risk level as the threshold. 
We further argue that the same risk level can be used here for all entity trust scores.
This is based on the fact that any risks posed by a resource will impact the risk level for all entities.
This is because an outdated system patch level can be exploited, for example, by a maliciously modified packet, a malicious user, but also by an infected device.

Analogously to the trust scores, the risk level can be calculated by an algorithm based on attributes with assigned weights.
Risk attributes are used for this purpose and their weights are evaluated analogously to the trust attribute weights. 
For risk attributes, $W_{wv}$ weights the risk of damage to an enterprise if the RAR were granted. 
The higher the risk inherent in a target value $W_{tv}$, the higher the mapping weight value in $W_{wv}$.
Below we provide a comprehensive list of risk attributes based on systematic literature review:

\begin{itemize}
    \item Request Protocol: Protocol used for the requested access, e.g. HTTP or FTP~\cite{DiDi20}.
    \item Request Action: Requested interaction with the resource, e.g. HTTP GET or POST~\cite{garbisEnterpriseGuide}.
    \item Data Sensitivity: Sensitivity of the requested data and the resulting potential damage in the event of a leak~\cite{chen2020security}.
    \item Service Software Patch Level: Patch level of the software serving the request~\cite{garbisEnterpriseGuide, GiBa17}.
    \item System State: Operational state of the system, e.g. normal or in maintenance~\cite{GiBa17}.
    \item System Threat Level: Security state of the system, e.g. normal or under attack~\cite{garbisEnterpriseGuide, GiBa17, RoBo20}.
    \item System Patch Level: Patch level of the system software, especially OS patch level~\cite{garbisEnterpriseGuide, GiBa17}.
    \item Network State: Operational state of the network, e.g. normal or in maintenance~\cite{aliyu2017trust}.
    \item Network Threat Level: Security state of the network, e.g. normal or under attack~\cite{garbisEnterpriseGuide, GiBa17, RoBo20}.
\end{itemize}

Two concrete algorithms for calculating the risk level including risk attributes with associated weights are introduced in Sections~\ref{sec:additiveTrustAlgorithms} and~\ref{sec:slTrustAlgorithm}.



%% file: chapters/06-score-based-trust-algorithms.tex
\section{Additive-based Trust Algorithm}
\label{sec:additiveTrustAlgorithms}
In this section, we introduce a straightforward additive trust algorithm.
It illustrates the principles from Sections~\ref{sec:threatModelTrustEntities},~\ref{sec:weightedAttributes}, and~\ref{sec:thresholdCalculation} and is later used to compare against the Subjective-Logic-based approach introduced in Section~\ref{sec:slTrustAlgorithm}.
For the sake of simplicity, a single total trust score combining all entities' trust attributes is calculated additively for each RAR.
This score is then compared to the RAR's risk level, and access is granted only if the trust score surpasses the risk level.

\subsection{Trust Score}
For each RAR, the total trust score $TS_{total} \in \mathbb R$ is initialized with 0.
After that, all considered trust attributes are evaluated.
$W_{tv}$ and $W_{wv} = \{wv | wv \in \mathbb R\}$ must be defined for each attribute where $f_{W} : W_{tv} \rightarrow W_{wv}$.
If $w_{av} \in W_{tv}$, the respective weight value $wv$ where $f_{W}(w_{av}) \rightarrow wv$ is added to $TS_{total}$. This results in the following equation:

\vspace{-4mm}
\begin{equation}
TS_{total} =
  \begin{cases}
    TS_{total} +  f_{W}(w_{av})     & \text{if } w_{av} \in W_{tv}\\
    TS_{total}  & \text{else}
  \end{cases}
\end{equation}

For example, Alice wants to access her enterprise's Gitlab. 
She uses \emph{Password Authentication} and her correct password is "1234". 
For her, this results in $W_{tv} = \{"1234"\}$.
We have chosen the example weight of 5 to add in case the entered password is correct.
Consequently, the set of weight values is $W_{wv} = \{5\}$ where $f_{W}("1234") \rightarrow 5$.
Alice enters her password correctly ($w_{av} = "1234"$), which results in $TS_{total} = 0 + f_{W}("1234") = 5$.


\subsection{Risk Level}
After $TS_{total}$ has been calculated, it is compared to a total risk level $RL_{total} \in \mathbb R$, which is initialized with 0. 
To determine the final $RL_{total}$, all risk attributes are evaluated analogously to the trust attributes.
$RL_{total}$ is then formed additively in the same way as $TS_{total}$:

\vspace{-4mm}
\begin{equation}
RL_{total} =
  \begin{cases}
    RL_{total} + f_{W}(w_{av})   & \text{if } w_{av} \in W_{tv}\\
    RL_{total}  & \text{else}
  \end{cases}
\end{equation}

For example, we consider again the enterprise's Gitlab service.
For the risk attribute \emph{System Patch Level}, the enterprise maintains two values: up-to-date and outdated. 
The enterprise considers access to up-to-date systems as not risky and thus up-to-date is not an element of the set of target values $W_{tv}$ for this risk attribute.
Consequently, $W_{tv} = \{\emph{"outdated"}\}$ and $W_{wv} = \{10\}$ where $f_{W}(\emph{"outdated"}) \rightarrow 10$. 
Assuming the Gitlab server runs an outdated version ($w_{av} = \emph{"outdated"}$), this leads to $RL_{total} = 0 + f_{W}(\emph{"outdated"}) = 10$.




\subsection{Access Decision}
Access to the service is granted only if the following inequality is satisfied:

\vspace{-4mm}
\begin{equation}
    TS_{total} > RL_{total}
\end{equation}

Based on the previous example for $RL_{total} = 10$ and $TS_{total} = 5$, this leads to the inequality $5 > 10$ and thus the RAR is rejected.

\section{Subjective Logic-based Trust Algorithm}
\label{sec:slTrustAlgorithm}

We now introduce a fine-grained trust algorithm using Subjective Logic (SL)~\cite{Au16}, contrasting the additive approach. The SL-based method considers individual entity trust scores and varying degrees of attribute fulfillment, along with dynamic risk levels. 

SL is used for decision-making under uncertainty. Uncertainty in a decision arises when a decision-maker lacks complete information at the time of making a decision. In the ZT context, SL informs access decisions based on trust, as for the additive approach. However, this approach offers finer granularity by forming separate opinions on the trustworthiness of the \emph{user}, \emph{device}, and \emph{communication channel}. Access is granted only if the trust scores for each entity surpass the associated risk level. We will outline how these individual trust scores and risk levels are calculated.

\subsection{User Trust Score}
To obtain the user trust score, we first form a binary opinion $\omega_X^A$ about whether the user is considered trustworthy.
In SL, $\omega_X^A$ represents the opinion about the proposition $X$ made by the subjective agent $A$. 
The proposition $X$ is a random variable $X \in \mathbb{X} = \{x, \overline{x}\}$.
In the case of the user trust, the binary random variable $UT \in \mathbb{X} = \{ut, \overline{ut}\}$ represents the proposition about the user trustworthiness \emph{UT}:

\begin{itemize}
    \item $ut:$ the user who performs this RAR is trustworthy
    \item $\overline{ut}:$ the user who performs this RAR is not trustworthy
\end{itemize}

Here, $A$ represents the PDP, which forms an opinion about the user trustworthiness.
This results in the opinion $\omega_{ut}^{\text{PDP}} = \{b_{ut}, d_{ut}, u_{ut}, a_{ut}\}$. 
The parameters are defined as follows:
\begin{itemize}
    \item $b_{ut}$: \emph{belief} that the user is trustworthy based on evidence
    \item $d_{ut}$: \emph{disbelief} that the user is trustworthy based on evidence
    \item $u_{ut}$: \emph{uncertainty} whether the user is trustworthy due to lack of evidence
    \item $a_{ut}$: \emph{base rate} represents prior knowledge about user trustworthiness. If no prior knowledge is available, SL defines a default value for binary opinions of 0.5.
\end{itemize}
 where $b_{ut}, d_{ut}, u_{ut}, a_{ut} \in [0,1]$ and the following additive requirement~\cite{Au16} is satified:

\vspace{-3.5mm}
\begin{equation}
    b + d + u = 1
\end{equation}
\vspace{-5mm}

The overall opinion $\omega_{ut}^{\text{PDP}}$ is a composition of individual opinions from subjective agents evaluating trust attributes. In our approach, we have on agent for each trust attribute. Each agent forms its own opinion $\omega_{ut}^{\text{Attribute}}$ based on the evidence of the attribute it evaluates. These opinions are then merged to create $\omega_{ut}^{\text{PDP}}$ using the weighted belief fusion operator $\WBF$ for SL, as defined in~\cite{DBLP:journals/corr/abs-1805-01388}. We argue that this fusion is well suited for trust-building as it allows for trust attribute weighting. For instance, retina authentication could be weighted higher than password authentication due to its arguably better security. This fusion express the weighting of opinions based on their uncertainty. Consequently, the set of weight values is defined as $W_{wv} = \{u^{A}_{ut} | u^{A}_{ut} \in [0,1]\}$. 
The lower the uncertainty of an opinion, the higher the weighting during the fusion.
The fused opinion is defined as $\omega_{ut}^{\text{PDP}} = (b^{\WBF \mathbb{A}}_{ut}, d^{\WBF \mathbb{A}}_{ut}, u^{\WBF \mathbb{A}}_{ut}, a^{\WBF \mathbb{A}}_{ut})$ where $\mathbb{A}$ is the set of subjective agents; in this case all agents that evaluate user trust attributes. From this overall opinion $\omega_{ut}^{\text{PDP}}$, we calculate the user trust score using projected probability~\cite{Au16} $P(UT = ut) = b^{\WBF \mathbb{A}}_{ut} + u^{\WBF \mathbb{A}}_{ut} \cdot a^{\WBF \mathbb{A}}_{ut}$.
This user trust score represents the probability that the user is trustworthy.

We now provide a simplified example using the subjective agent~\emph{PWAuth}, which forms an opinion $\omega_{ut}^{\text{PWAuth}} = \{b_{ut}, d_{ut}, u_{ut}, a_{ut}\}$ on user trustworthiness. This opinion is based on two evidence factors: The correctness of the entered password and the already failed attempts to enter the correct password. So here $w_{av}$ is a pair $(pw, att)$ where $pw$ is the entered password and $att  \in \mathbb N_0$ is the amount of already failed attempts. 
Consequently, $W_{tv}$ is a set of pairs $(pwc, att)$ where $pwc$ is the correct password.
Assuming Alice enters her password correctly after 5 failed attempts ($w_{av} = (pwc, 5)$), this results in the following example parameter values:
\begin{itemize}
    \item $b_{ut}$ = 0.2 (evidence that user is trustworthy as $w_{av} \in W_{tv}$)
    \item $d_{ut}$ = 0.6 (evidence for untrustworthiness as $att \neq 0$, which could be a sign of a password guessing attack)
    \item $u_{ut}$ = 0.2 (A residue of uncertainty since Alice could have mistyped 5 times)
    \item $a_{ut}$ = 0.5 (no prior knowledge is available)
\end{itemize}
More failed attempts reduce uncertainty $u_{ut}$ as they serve as evidence, making the subjective agent \emph{PWAuth} more confident in its opinion. This decrease in uncertainty leads to increased disbelief $d_{ut}$, as accumulating failed attempts suggest an illegitimate login attempt.
For the attribute weights, we get the mapping $f_{W}: (pwc, att) \rightarrow u_{ut}$.
For Alice it is $f_{W}: (pwc, 5) \rightarrow 0.3$.
Since we are evaluating only one attribute in this example, we obtain the overall opinion $\omega_{ut}^{\text{PDP}} = \omega_{ut}^{\text{PWAuth}} = \{0.2, 0.6, 0.2, 0.5\}$ from it. We can now calculate the probability $P(UT = ut) = 0.2 + 0.2 \cdot 0.5 = 0.3$ that Alice is trustworthy.

\subsection{Device Trust Score}
For device trustworthiness \emph{DT}, the random variable $\emph{DT} \in \mathbb{X} = \{dt, \overline{dt}\}$ has the two outcomes:

\begin{itemize}
\item $dt :$ the device that performs this RAR is trustworthy
\item $\overline{dt} :$ the device that performs this RAR is not trustworthy
\end{itemize}

This results in the overall opinion $\omega_{dt}^{\text{PDP}} = (b^{\WBF \mathbb{A}}_{dt}, d^{\WBF \mathbb{A}}_{dt}, u^{\WBF \mathbb{A}}_{dt}, a^{\WBF \mathbb{A}}_{dt})$, which is build analogously as for the user trustworthiness. 
From this opinion, the device trust score $P(DT = dt)  = b^{\WBF \mathbb{A}}_{dt} + u^{\WBF \mathbb{A}}_{dt} \cdot a^{\WBF \mathbb{A}}_{dt}$ is calculated.

\subsection{Communication Channel Trust Score}
Here, the PDP forms the overall opinion $\omega_{cct}^{\text{PDP}} = (b^{\WBF \mathbb{A}}_{cct}, d^{\WBF \mathbb{A}}_{cct}, u^{\WBF \mathbb{A}}_{cct}, a^{\WBF \mathbb{A}}_{cct})$ about the communication channel trustworthiness $\emph{CCT} \in \mathbb{X} = \{cct, \overline{cct}\}$ with the two outcomes:

\begin{itemize}
\item $cct :$ the communication channel is trustworthy
\item $\overline{cct} :$ the communication channel is not trustworthy
\end{itemize}

The overall opinion $\omega_{cct}^{\text{PDP}}$ is formed analogously as for the user trustworthiness.
From this, the communication channel trust score $P(CCT = cct) = b^{\WBF \mathbb{A}}_{cct} + u^{\WBF \mathbb{A}}_{cct} \cdot a^{\WBF \mathbb{A}}_{cct}$ is calculated.

\subsection{Risk Level}
\label{sssec:risk}

For determining the risk level, an opinion $\omega_{rod}^{\text{PDP}}$ is build on the risk of damage (ROD) to the enterprise if the RAR is permitted.
This proposition is represented by the random variable $\text{ROD} \in \mathbb{X} = \{rod, \overline{rod}\}$ with

\begin{itemize}
    \item $rod :$ Permitting the RAR will cause damage
    \item $\overline{rod} :$ Permitting the RAR will not cause damage
\end{itemize}

The parameters for the risk opinion $\omega_{rod}^{\text{PDP}} = \{b_{rod}, d_{rod}, u_{rod}, a_{rod}\}$ are defined as
\begin{itemize}
    \item $b_{rod}$ : \emph{belief} that permitting the RAR causes damage to the enterprise based on evidence
    \item $d_{rod}$ : \emph{disbelief} that permitting the RAR causes damage to the enterprise based on evidence
    \item $u_{rod}$ : \emph{uncertainty} whether permitting the RAR causes damage to the enterprise due to lack of evidence
    \item $a_{rod}$ : \emph{base rate} based on prior knowledge about permitted RARs that caused damage to the enterprise
\end{itemize}
 where $b_{rod}, d_{rod}, u_{rod}, a_{rod} \in [0,1]$ and the additive requirement (4) is satisfied.

The risk opinion $\omega_{rod}^{\text{PDP}}$ is a composition of the opinions formed by the respective subjective agents.
Each subjective agent collects evidence based on the respective risk attribute it evaluates.
All individual opinions regarding risk are then fused to obtain the overall opinion $\omega_{rod}^{\text{PDP}}$. 
As fusion we use the associative cumulative fusion operator $\diamond$ defined in~\cite{Au16}. 
The cumulative fusion is defined as $\omega_{X}^{\text{A}} = (b^{A \diamond B}_{X}, d^{A \diamond B}_{X}, u^{A \diamond B}_{X}, a^{A \diamond B}_{X})$ where $A$ and $B$ are two agents that formed an opinion about the same proposition $X$.
For the total risk opinion $\omega_{rod}^{\text{PDP}}$ we fuse all single agents' opinions sequentially.
This results in the overall opinion $\omega_{rod}^{\text{PDP}} = (b^{\diamond \mathbb{A}}_{rod}, d^{\diamond \mathbb{A}}_{rod}, u^{\diamond \mathbb{A}}_{rod}, a^{\diamond \mathbb{A}}_{rod})$ where ${\diamond \mathbb{A}}$ represents the set $\mathbb{A}$ of all risk agents that are all fused by the operator $\diamond$.
We argue that the fusion operator used for trust scores is unsuitable here, as weighted fusion can lead to lower belief values (here: the belief in causing damage). Unlike trust scores, where reduced belief in trustworthiness is acceptable if single trust attributes are not met, the belief in the risk of damage should not decrease. For instance, an outdated system patch can not be compensated by an updated service patch. Existing risks just add up. A cumulative fusion method, summing the evidence parameters b and d from individual risk opinions, is more appropriate here. Consequently, for the SL risk level calculation risk attribute weights are directly derived from the respective opinion's evidence parameters $b^{A}_{rod}$ and $d^{A}_{rod}$: $W_{wv} = \{(b^{A}_{rod}, d^{A}_{rod}) | b^{A}_{rod}, d^{A}_{rod} \in [0,1]\}$. To get the final risk level, the projected probability $P(\text{ROD} = rod) = b_{rod}^{\diamond \mathbb{A}} + u_{rod}^{\diamond \mathbb{A}} \cdot a_{rod}^{\diamond \mathbb{A}}$ is formed.

As an example, this could result in a risk opinion $\omega_{rod}^{\text{SPL}} = \{b_{rod}, d_{rod}, u_{rod}, a_{rod}\}$ formed by the agent that evaluates the attribute \emph{System Patch Level} of the Gitlab server.
The enterprise manages two patch levels: \emph{"outdated"} and \emph{"up-to-date"}.
Thus, the set of attribute target values is $W_{tv} = \{\emph{"outdated"}, \emph{"up-to-date"}\}$.
The example parameter values for a system with the latest patch ($w_{av} = \emph{"up-to-date"}$) could look like this:

\begin{itemize}
    \item $b_{rod}$ = 0.0 (no evidence that permitting this RAR will cause damage to the enterprise)
    \item $d_{rod}$ = 0.8 (evidence that permitting this RAR will cause no damage as $w_{av} = \emph{"up-to-date"}$)
    \item $u_{rod}$ = 0.2 (uncertainty since there could be undetected vulnerabilities for the latest patch)
    \item $a_{rod}$ = 0.5 (no prior knowledge is available)
\end{itemize}
For the function $f_{W}: W_{tv} \rightarrow (b^{A}_{rod}, d^{A}_{rod})$, we then get the example mapping $f_{W}: \emph{"up-to-date"} \rightarrow (0.0, 0.8)$.
Since we consider only one attribute in the example, we get the overall risk opinion $\omega_{rod}^{\text{PDP}} = \omega_{rod}^{\text{SPL}} = \{0.0, 0.8, 0.2, 0.5\}$. 
This results in $P(\text{ROD} = rod) = 0.0 + 0.2 \cdot 0.5 = 0.1$.


\subsection{Access Decision}
To make an access decision, the individual entity trust scores are compared to the risk level.
As we will discuss in Section~\ref{sec:discussion}, comparing individual scores to the risk level prevents trust compensation.
Access is only permitted if the trust in each entity exceeds the risk of damage. 
This results in the following three inequalities that must all be satisfied:
\vspace{-1mm}
\begin{equation}
    P(ut) > P(rod), 
    P(dt) > P(rod),
    P(cct) > P(rod)
\end{equation}
\vspace{-5mm}

The example user trust score ($P(ut) = 0.3$) and risk level ($P(rod) = 0.1$) lead to the inequality $0.3 > 0.1$.
Considered for the user trust score, the RAR would be permitted.

%% file: chapters/07-evaluation-discussion.tex
\section{Discussion \& Related Work}
\label{sec:discussion}
Following the introduction of both trust algorithms, we explore their qualitative aspects, such as expressiveness.

Score-based trust algorithms use attribute weights for trust score calculations. In the additive method, an attribute is either fully weighted or not, while the SL-based approach allows for nuanced weighting through continuous belief and disbelief values. For example, as discussed in Section~\ref{sec:slTrustAlgorithm}, more failed password attempts by Alice increases disbelief in her identity. As we gather evidence through failed attempts, the uncertainty diminishes, leading to a higher weighting of the attribute. While prior studies have explored trust score calculations, they often have a different focus or less generality. Dimitrakos et al.~\cite{DiDi20} implemented an SL-based approach with focus on device authentication by sensors in the IoT domain.
The authors in~\cite{da2021zero} use different discrete values for specifying their trust weights in the context of smart homes.
Albuali et al.~\cite{albuali2020ztimm} provide a fine-granular model while focusing on three behavioral aspects in cloud computing.
In~\cite{tao2018fine} and~\cite{channel} fixed weights are considered for their calculations.
Yao~\cite{yao2020dynamic} calculate attribute target values for three behavioral aspects using historical data but have not yet elaborated on their method for setting and adjusting weights.
While our SL-based algorithm allows for fine-grained weight adjustments, it increases the demand to determine reasonable weights by a security expert for each use case.
Additionally, the mathematical complexity of the SL-based approach increases the risk of errors.

Similarly, we have explored different trust score approaches in our algorithms. The additive approach calculates a single, total trust score, a common practice in the majority of the literature such as in~\cite{yao2020dynamic, mandal2021cloud, ghate2021advanced} or~\cite{da2021zero}. In~\cite{channel}, single trust scores for the user, terminal, and channel are calculated but before being compared to the threshold, all are subtracted from each other to build one total score. One total trust score simplifies risk-level reconciliation but lacks granularity. In contrast, our SL-based approach calculates separate scores for each entity, allowing for finer access control and preventing unintended trust compensation. For instance, a high device score should not offset an untrustworthy user. 
However, managing individual scores for each entity complicates matching them with appropriate risk levels.

In addition to entity trust scores, we presented a calculation method for the threshold, we refer to as risk level.
As discussed in Section~\ref{sec:thresholdCalculation}, this threshold is a relatively static value in the current literature, with mostly no guidance on how to set it.
By calculating the risk level based on risk attributes, it is possible to dynamically recalculate this threshold for each RAR.
Due to the weighted influence of the individual risk attributes, the adjustment can be realized with the same granularity as for the trust value.
This also means that the threshold can be automatically adapted to the enterprise's current risk situation for each RAR.

\section{Performance Evaluation}
\label{sec:evaluation}
As poor user experience can compromise security, performance is a crucial factor~\cite{GiBa17}. In this section, we evaluate the decision-making time of both trust algorithms in a realistic ZT enterprise setting following ZT requirements of mutual authentication, encryption, and integrity protection for all communication channels~\cite{BrKa21}. We aim to assess (1) how the fine-grained SL-based algorithm impacts decision-making time compared to the straightforward additive algorithm, and (2) the effect of dynamic versus static risk level calculation, the latter commonly seen in the literature.

We set up a testbed featuring all components shown in Figure~\ref{fig:zt_background}. Each of the four servers, equipped with an Intel Xeon E5-2630 v3 CPU (32 Cores, 64 Threads), 128 GB DDR4 RAM, and a Samsung SSD 850 EVO 500GB, are interconnected using Mellanox Connect-X4 100G network cards and direct server cabling. Implemented in Golang 1.20.2 on Ubuntu 22.04.2, the setup aligns with Section~\ref{sec:background} and is publicly available on GitHub\footnote{\url{https://github.com/zintpavowj/ZT_Score-based_Network_Level_AC}}. All communications are secured with mutual TLS. The databases and ZT components communicate via REST APIs and SQL, respectively. We filled all 29 trust and 9 risk attributes with dummy values and weights, which do not affect performance compared to realistic data.

We measured the decision-making time from when the PEP sent the decision request to when the authorization decision was received, as shown in Figure~\ref{fig:zt_background}. That allowed us to assess the impact of trust algorithm differences, also considering other factors like network operations. Both trust algorithms’ pure calculation time was in the low microsecond range (10-40$\mu$s). We measured decision-making time in two scenarios, comparing additive and SL-based approaches with both static and dynamic risk levels. The PEP ran 1 to 128 parallel instances, each sending 1000 requests. Tests were repeated 10 times and reported values represent the median.

\begin{figure}[h]
    \centering
    \includegraphics[width=.85\columnwidth]{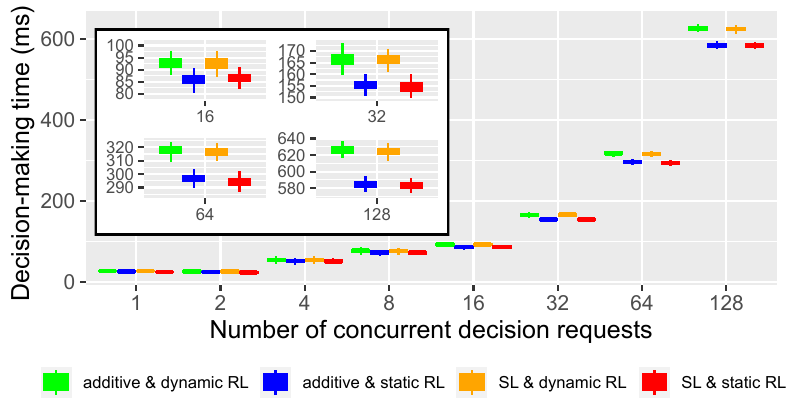}
    \vspace{-3.5mm}
    \caption{Decision-making time in the case the PIP queries all attributes from the database. For the measured times, the quartiles are presented.}
    \label{fig:resultsWoCache}
\end{figure}

\begin{figure}[h]
    \centering
    \includegraphics[width=.85\columnwidth]{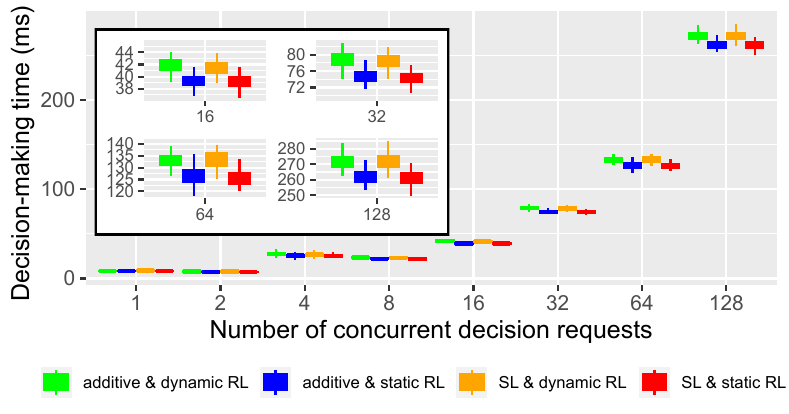}
    \vspace{-3.5mm}
    \caption{Decision-making time in the case the PIP holds all attributes in the cache. For the measured times, the quartiles are presented.}
    \label{fig:resultsWithCache}
\end{figure}

In the first test scenario, we examined the PIP's worst-case scenario, where its attribute cache is empty and all trust attributes must be retrieved from the database. 
As shown in Figure~\ref{fig:resultsWoCache}, decision-making time increases even before reaching the optimal CPU load of 64 parallel PEP instances due to blocking network system calls in kernel space. 
At 64 parallel PEP instances, both the additive trust algorithm (296ms) and the SL-based trust algorithm (294ms) exhibit nearly identical decision-making times with a static risk level. However, when calculating a dynamic risk level, which requires loading additional risk attributes from Section~\ref{sec:thresholdCalculation}, decision-making time increases by 7\% for both algorithms (317ms for additive, 316ms for SL-based). This slight difference is due to all attributes being loaded in one batch per request.
\vspace{-2.5mm}
\vspace{-1mm}
Once the PIP has loaded all attributes, it serves the PDP's attribute requests from its cache.
The results of this second test scenario are shown in Figure~\ref{fig:resultsWithCache}.
In the case of 64 parallel PEP instances, the decision-making time for both trust algorithms with static risk level (126ms) as well as with dynamic risk level calculation (133ms) is about 42\% shorter compared to the first test scenario.
\vspace{-2.5mm}
\vspace{-1mm}
The tests showed that both trust algorithms have the same decision-making time. 
The additional dynamic risk level calculation increases the decision-making time by only 7\% and is of little impact compared to the dominant time factors of database transfers and network-related operations. In summary, dynamic risk level calculation and the SL-based trust algorithm do not have a significant impact on performance.

%% file: chapters/08-conclusion.tex
\section{Conclusion \& Future Work}
\label{sec:conclAndFutureWork}
In this paper, we initially identify three key trust entities - \emph{user}, \emph{device}, and \emph{communication channel} - essential in RARs. These are motivated by the ZT threat model, which presumes initial distrust in these entities. Hence, trust calculations for RARs must include all three.

Trust in the three entities is determined through trust attributes. Based on a systematic literature review, we provide a comprehensive list of 29 such trust attributes for enterprise settings, categorized by applicable entities. Due to a gap in the literature for attribute weighting, we introduce a mathematical model for this purpose. Our model maps continuous trust-building target values to weights and allows for varying degrees of attribute fulfillment.

For access decisions, calculated trust scores are compared to a threshold. We argued that the required trust level depends on the potential risk of damage from granting access, leading us to introduce a novel risk level calculation method as the threshold. Similar to trust scores, this risk level is determined by weighted risk attributes, enabling dynamic risk level calculations for each RAR.

To illustrate the discussed contributions, we first described a basic additive trust algorithm that calculates a single trust score, compared against a dynamically calculated risk level. We then introduced a new trust algorithm based on SL, which effectively utilizes all prior contributions. This algorithm compares separate trust scores for each entity, allowing for nuanced attribute fulfillment, against a risk level.

We observed that the SL-based algorithm allows for fine-grained trust calculations, improving decision expressiveness but adding complexity in attribute weighting. The performance evaluation showed that this complexity, along with dynamic risk levels, does not significantly impact decision-making time compared to the straightforward additive approach with a static risk level. Therefore, performance is not a barrier to using the more expressive SL-based approach. Future research will explore if this leads to more accurate access decisions.

Therefore, the next step is to evaluated how the weights for the attributes can be determined best for each trust algorithm.
Currently the usual approach is that for each use case the responsible security expert has to define them.
Based on this, it is necessary to investigate which approach leads to more accurate access decisions. 
Here, accurate is to be understood as the lowest possible false positive and false negative rates.